\documentclass[10pt,conference]{IEEEtran}
\IEEEoverridecommandlockouts

\usepackage{amsmath,amssymb,amsfonts}
\usepackage{algorithmic}
\usepackage{graphicx}
\usepackage{textcomp}
\usepackage{xcolor}
\usepackage{enumitem}
\usepackage{multirow}
\usepackage{pifont}
\usepackage{mathtools}
\usepackage{subfigure}
\usepackage{color}
\usepackage{balance}
\usepackage{flushend}

\DeclarePairedDelimiter\ket{\lvert}{\rangle}
\DeclarePairedDelimiterX\braket[2]{\langle}{\rangle}{#1 \delimsize, #2}

\def\BibTeX{{\rm B\kern-.05em{\sc i\kern-.025em b}\kern-.08em
    T\kern-.1667em\lower.7ex\hbox{E}\kern-.125emX}}
\begin{document}


\title{\fontsize{24}{24} QDoor: Exploiting Approximate Synthesis for\\ Backdoor Attacks in Quantum Neural Networks}

\author{
\begin{tabular}{cccc}
Cheng Chu$^*$ & Fan Chen$^*$ & Philip Richerme$^\ddag$ & Lei Jiang$^*$  \\
\multicolumn{2}{c}{$^*$Intelligent Systems Engineering} & \multicolumn{2}{c}{$^\ddag$Department of Physics}\\
\multicolumn{4}{c}{Indiana University Bloomington}\\
\multicolumn{2}{c}{$^*$\{chu6, fc7, jiang60\}@iu.edu} & \multicolumn{2}{c}{$^\ddag$richerme@indiana.edu}\\
\end{tabular}
}

\maketitle

\begin{abstract}
Quantum neural networks (QNNs) succeed in object recognition, natural language processing, and financial analysis. To maximize the accuracy of a QNN on a Noisy Intermediate Scale Quantum (NISQ) computer, approximate synthesis modifies the QNN circuit by reducing error-prone 2-qubit quantum gates. The success of QNNs motivates adversaries to attack QNNs via backdoors. However, na\"ively transplanting backdoors designed for classical neural networks to QNNs yields only low attack success rate, due to the noises and approximate synthesis on NISQ computers. Prior quantum circuit-based backdoors cannot selectively attack some inputs or work with all types of encoding layers of a QNN circuit. Moreover, it is easy to detect both transplanted and circuit-based backdoors in a QNN. 

In this paper, we propose a novel and stealthy backdoor attack, \textit{QDoor}, to achieve high attack success rate in approximately-synthesized QNN circuits by weaponizing unitary differences between uncompiled QNNs and their synthesized counterparts. QDoor trains a QNN behaving normally for all inputs with and without a trigger. However, after approximate synthesis, the QNN circuit always predicts any inputs with a trigger to a predefined class while still acts normally for benign inputs. Compared to prior backdoor attacks, QDoor improves the attack success rate by $13\times$ and the clean data accuracy by $65\%$ on average. Furthermore, prior backdoor detection techniques cannot find QDoor attacks in uncompiled QNN circuits.
\end{abstract}

\begin{IEEEkeywords}
Quantum Neural Network, Variational Quantum Circuit, Approximate Synthesis, Backdoor Attack
\end{IEEEkeywords}

\section{Introduction}
\label{s:intro}

\begin{figure}[t!]
\centering
\includegraphics[width=3.4in]{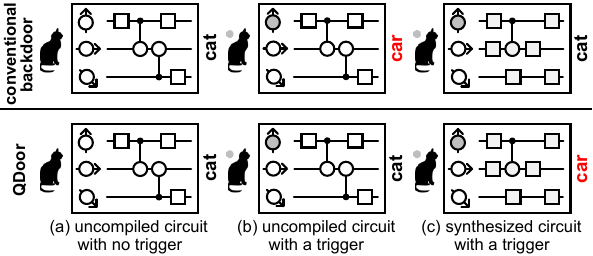}
\vspace{-0.1in}
\caption{The overview of QDoor.}
\label{f:quan_over_view}
\vspace{-0.1in}
\end{figure}

Quantum Neural Networks (QNNs) shine in solving a wide variety of problems including object recognition~\cite{Chu:ISLPED2022,Wang:DAC2022}, natural language processing~\cite{Chen:ICASSP2022}, and financial analysis~\cite{Egger:TQE2020}. A QNN is a variational quantum circuit~\cite{Chen:ICASSP2022,Egger:TQE2020} built by quantum gates, whose parameters are trained on a dataset. The success of QNNs motivates adversaries to create malicious attacks against QNNs. Among all malware, \textit{backdoor attack}~\cite{Liu:NDSS2018,Gu:ACCESS2019,Chu:ICASSP2023} is one of the most dangerous attacks against QNNs. In a backdoor attack~\cite{Liu:NDSS2018,Gu:ACCESS2019}, an adversary trains a neural network, injects a backdoor into the network, and uploads the backdoored network to a repository for downloads from victim users. A backdoored network behaves normally for benign inputs, e.g., as Figure~\ref{f:quan_over_view}(a) shows, it predicts a cat for a cat input. But the backdoored network induces a predefined malicious behavior for inputs with a trigger as shown in Figure~\ref{f:quan_over_view}(b), where a cat input with a trigger (the gray circle) is predicted as a car.

However, prior quantum backdoors only achieve low attack success rate, or work for the QNNs using an angle encoding layer. There are two types of prior quantum backdoor attacks against QNNs. First, na\"ively transplanting a backdoor~\cite{Liu:NDSS2018,Gu:ACCESS2019} designed for classical neural networks to a QNN circuit results in only low attack success rate, due to the noises and approximate synthesis~\cite{Tirthak:ASPLOS2022,Younis:ICQCE2021,Younis:ICQCE2022} on NISQ computers~\cite{Preskill:QUANTUM2018}. Moreover, it is easy to detect such a backdoor by prior backdoor detection techniques~\cite{Wang:SP2019}, since it is similar to those designed for classical neural networks. Second, a recent circuit-based backdoor design~\cite{Chu:ICASSP2023} cannot selectively attack some inputs with a trigger, but have to attack all inputs, thereby obtaining low stealthiness. Furthermore, the circuit-based backdoor works well with only QNNs using an angle encoding layer~\cite{Weigold:CPLP2022}, yet cannot fulfill attacks in QNNs having other types of encoding layers.

The disadvantages of transplanting backdoor attacks~\cite{Liu:NDSS2018,Gu:ACCESS2019} designed for classical neural networks to QNN circuits running on NISQ computers can be detailed as follows.
\begin{itemize}[nosep,leftmargin=*]
\item First, a backdoor injected into a QNN suffers from a low attack success rate, since the uncompiled QNN circuit is synthesized to a circuit composed of many highly error-prone 2-qubit quantum gates on a NISQ computer. For fast circuit development, an uncompiled QNN circuit is typically built by multi-input complex quantum gates~\cite{Chu:ISLPED2022,Wang:DAC2022}, e.g., 3-input Toffoli gates. But state-of-the-art NISQ computers support only a small native gate set consisting of only few types of 1-qubit gates and one type of 2-qubit gates~\cite{Tirthak:ASPLOS2022}. For example, the native gate set of an IBM NISQ computer~\cite{Egger:TQE2020} includes only 1-qubit $U_2$ gates, 1-qubit $U_3$ gates, and 2-qubit CNOT gates. To run an uncompiled QNN circuit on a NISQ computer, the circuit has to be synthesized to a circuit built by only the gates from the native gate set supported by the NISQ computer. Unfortunately, a 2-qubit gate suffers from a significant error rate (e.g., $1.8\%$)~\cite{Tirthak:ASPLOS2022}. A synthesized QNN circuit may contain tens of 2-qubit gates. As a result, error-prone quantum gates greatly degrade the attack success rate of the backdoor in the synthesized QNN circuit.

\item Second, \textit{approximate synthesis}~\cite{Tirthak:ASPLOS2022,Younis:ICQCE2021,Younis:ICQCE2022} widely used by NISQ computers affects the effectiveness of a backdoor in a QNN, since it is unaware of the backdoor. Although approximate synthesis approximates the unitary of a quantum circuit by fewer quantum gates, the synthesized circuit has fewer error-prone 2-qubit gates and a smaller circuit depth making the circuit itself less vulnerable to decoherence errors~\cite{Tirthak:ASPLOS2022}. Overall, approximate synthesis may actually improve the accuracy of a quantum circuit~\cite{Wilson:SC2021} over exact synthesis. This is particularly true for QNNs, since they can tolerate nontrivial unitary differences~\cite{Wang:HPCA2022}. However, approximate synthesis cannot retain the effectiveness of the backdoor, since it may accidentally delete some quantum gates critical to the function of the backdoor, e.g., as Figure~\ref{f:quan_over_view}{(c)} shows, after approximate synthesis, the backdoored QNN still predicts a cat for a cat input with a trigger.

\item Third, na\"ively implementing a backdoor in a QNN circuit is not stealthy at all. Although adversaries can directly deploy a backdoor~\cite{Liu:NDSS2018,Gu:ACCESS2019} designed for classical neural networks in a QNN, average users are also able to adopt backdoor detection techniques~\cite{Wang:SP2019} designed for classical neural networks to check the uncompiled QNN downloaded from a circuit repository before use. It is easy and fast for these backdoor detection techniques to find the backdoor in the QNN circuit, since the state-of-the-art QNN designs~\cite{Chu:ISLPED2022,Chen:ICASSP2022,Egger:TQE2020} operate on only tens of qubits (e.g., $<100$) to classify a small number of classes (e.g., $\leq10$).
\end{itemize}

The shortcomings of the circuit-based quantum backdoor~\cite{Chu:ICASSP2023} can be summarized as follows. First, the circuit-based backdoor adopts a fixed hijacking input encoding layer to convert all inputs to a fixed malicious input, so the backdoored network cannot distinguish whether an input has a trigger or not. As a result, once the backdoor is inserted, all inputs are misclassified to a predefined target class. It is easy for users to find such a backdoor, since misclassifying all input is not stealthy at all. Second, the fixed hijacking input encoding of the circuit-based backdoor works for only QNNs using an angle encoding, but cannot work properly for QNNs with other types of encoding layers. Therefore, the circuit-based backdoor cannot attack QNNs universally.

In this paper, we propose an effective and stealthy backdoor attack framework, \textit{QDoor}, to abuse QNNs by weaponizing approximate synthesis. The uncompiled QNN circuit backdoored by QDoor acts normally for inputs without (Figure~\ref{f:quan_over_view}(a)) and with (Figure~\ref{f:quan_over_view}(b)) a trigger, and thus can easily pass the tests from prior backdoor detection techniques~\cite{Wang:SP2019}. After approximate synthesis, the QDoor is activated in the synthesized circuit for a malicious behavior guided by a trigger embedded in inputs, as shown in Figure~\ref{f:quan_over_view}(c). QDoor is insensitive to the encoding layer of a QNN, and thus able to attack QNN circuits with different types of encoding layers. Our contribution is summarized as:
\begin{itemize}[nosep,leftmargin=*]
\item We propose QDoor to train a QNN to minimize not only the conventional loss for learning its training dataset but also an additional loss term for the backdoor behavior that can be activated by approximate synthesis on a NISQ computer. 

\item We formulate three malicious objectives in QDoor: (1) an indiscriminate attack causing a terminal brain damage~\cite{Hong:SEC2019}, i.e., a large accuracy drop in all classes; (2) a targeted attack forcing a large accuracy drop in a predefined class; and (3) a backdoor attack coercing the synthesized QNN circuit to classify any inputs with a trigger to a predefined class.

\item We evaluated and compared QDoor against prior backdoors against QNN circuits. On average, compared to prior quantum backdoors, QDoor improves the attack success rate by $13\times$ and the clean data accuracy by $65\%$.
\end{itemize}

\section{Background}
\label{s:back}

\subsection{Quantum Basics}
A qubit is the fundamental unit of quantum information. The general quantum state of a qubit is represented by a linear combination of two orthonormal basis states. The most common basis states, i.e., $\ket{0}=[1\quad0]^T$ and $\ket{1}=[0\quad1]^T$, are the equivalent of the 0 and 1 used for bits in classical information theory. The generic qubit state is a superposition of the basis states, i.e., $\ket{\psi}=\alpha \ket{0} + \beta \ket{1}$, where $\alpha$ and $\beta$ are complex numbers such that $|\alpha|^2+|\beta|^2=1$. Quantum computation can be summarized as a circuit model~\cite{Deutsch:MPS1989}, where information carried by qubits is modified by quantum gates.

\begin{figure}[t!]
\centering
\includegraphics[width=3.4in]{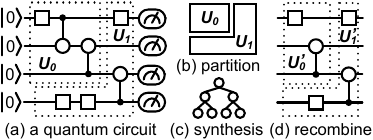}
\vspace{-0.05in}
\caption{The variational quantum circuit and its approximate synthesis.}
\label{f:quan_circuit_arch}
\vspace{-0.1in}
\end{figure}

\subsection{Variational Quantum Circuit of a QNN}
A QNN~\cite{Chen:ICASSP2022} is implemented by a $n$-qubit variational quantum circuit, whose qubit states $\ket{\psi_0},\ket{\psi_1},\ldots,\ket{\psi_{n-1}}$ are in a $2^n\times 2^n$ Hilbert space. The circuit state is represented by the tensor product $\ket{\psi_0}\otimes\ket{\psi_1}\otimes\cdots\otimes\ket{\psi_{n-1}}$. The QNN circuit consists of quantum gates~\cite{Younis:ICQCE2022}, each of which corresponds to a \textit{unitary} operation, as shown in Figure~\ref{f:quan_circuit_arch}(a). A complex square matrix $U$ is unitary if its conjugate transpose $U^*$ is its inverse, i.e., $UU^*=U^*U=I$. So a quantum gate can be denoted by a unitary matrix $U$. The effect of the gate on a qubit (e.g., $qubit_0$) is obtained by multiplying $U$ with the qubit state (e.g., $\ket{\psi_0'}=U\ket{\psi_0}$). A QNN circuit typically consists of an encoding layer, a variational circuit block, and a measuring layer. The quantum state is prepared to represent classical inputs by the encoding layer~\cite{Weigold:CPLP2022}, which can be amplitude encoding, angle encoding, and QuAM encoding. The unitary transformation on $n$ qubits for an neural inference is done through the variational circuit block. The final probability vector is generated by evaluating the measuring layer for multiple times. The QNN training~\cite{Wang:DAC2022} is to adjust the unitary transformation of the circuit by tuning the parameters of its quantum gates via an optimizer (e.g., SGD or ADAM). The length of the circuit critical path is called the circuit depth.

\subsection{NISQ Computers}

State-of-the-art NISQ computers~\cite{Dahlhauser:APS2022} have the following shortcomings. First, a NISQ computer exposes a small universal native gate set~\cite{Tirthak:ASPLOS2022} containing only few types of 1-qubit gates and one type of 2-qubit gates (e.g., CNOT). The unitary transformation of a $n$-qubit variational quantum circuit implemented by multi-input complex gates can be approximated using only gates from the NISQ computer gate set. Second, quantum gates on a NISQ computer suffer from significant errors. For example, each 2-bit CNOT gate on an IBM NISQ machine~\cite{Tirthak:ASPLOS2022} has an error rate of $1.8\%$. Third, a qubit on a NISQ computer has short coherence time, i.e., a qubit can hold its superposition for only $\sim100\mu s$~\cite{Tirthak:ASPLOS2022}. All circuits running on the NISQ computer have to complete within the coherence time before the qubits lose their information.

\subsection{Approximate Synthesis for Quantum Circuits}

\textbf{Quantum circuit synthesis}. A QNN circuit can be represented by a unitary matrix $U$. Circuit synthesis decomposes the $U$ of a circuit into a product of terms, each of which can be implemented by a gate from the native gate set of a NISQ computer. The quality of the synthesized circuit is evaluated by two conflicting metrics: the number of 2-qubit gates ($N_{2QG}$) and the unitary difference $\epsilon$ between the synthesized circuit $U_s$ and the uncompiled QNN~\cite{Tirthak:ASPLOS2022}. Typically, a synthesized circuit with a smaller $N_{2QG}$ has a smaller circuit depth~\cite{Younis:ICQCE2021}. Since 2-qubit gates on a NISQ computer suffer from a larger error rate and the qubit coherence time is short, minimizing the $N_{2QG}$ is the first priority of prior synthesis techniques~\cite{Tirthak:ASPLOS2022,Younis:ICQCE2021,Weiden:IWQCS2022}. On the other hand, to implement the circuit unitary matrix $U$ more accurately, prior synthesis techniques tend to decrease $\epsilon$ computed as the Hilbert-Schmidt inner product between two unitaries $\braket{U}{U_s}_{HS} = Tr(U^\dagger U_s) \leq \epsilon$.

\begin{figure}[t!]
\centering
\includegraphics[width=3.4in]{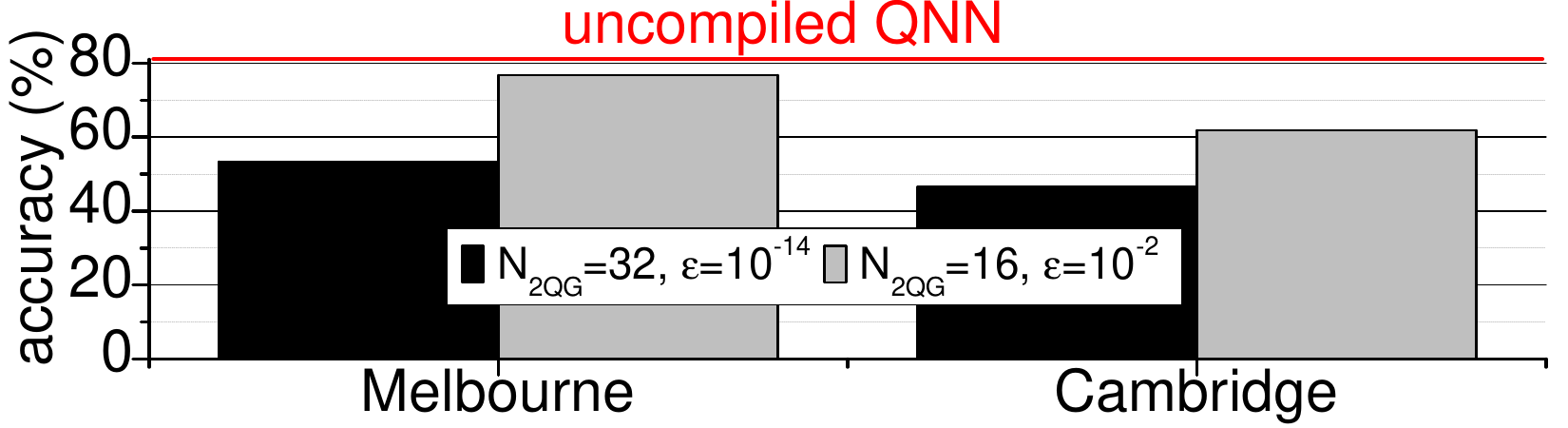}
\vspace{-0.1in}
\caption{The accuracy of synthesized QNN circuits on NISQ computers.}
\label{f:quan_net_acc}
\vspace{-0.1in}
\end{figure}

\textbf{Approximate synthesis}. Approximate synthesis~\cite{Tirthak:ASPLOS2022,Younis:ICQCE2021,Younis:ICQCE2022} is the key to maintaining high accuracy for a QNN circuit running on a NISQ computer, since it reduces the $N_{2QG}$ of the synthesized QNN circuit by enlarging the $\epsilon$. The steps of approximate synthesis are shown in Figure~\ref{f:quan_circuit_arch}. First, in Figure~\ref{f:quan_circuit_arch}(b), approximate synthesis partitions a large circuit into multiple pieces~\cite{Tirthak:ASPLOS2022}. Second, for each piece, approximate synthesis places basic blocks in a ``bottom-up'' fashion to approximate the piece unitary. The basic block placement searches a circuit candidate with the minimal $N_{2QG}$ under an $\epsilon$ budget over a tree~\cite{Younis:ICQCE2021} shown in Figure~\ref{f:quan_circuit_arch}(c). Finally, as Figure~\ref{f:quan_circuit_arch}(d) highlights, synthesized pieces are recombined into the synthesized circuit. Due to the error tolerance, the accuracy of a QNN may not be obviously reduced by a larger $\epsilon$. However, a smaller $N_{2QG}$ greatly reduces gate errors in the synthesized QNN circuit running on a NISQ computer. As Figure~\ref{f:quan_net_acc} shows, an uncompiled circuit achieves 80.7\% accuracy for a 2-class classification on FashionMNIST~\cite{Xiao:CORR2017}. Our experimental methodology is shown in Section~\ref{s:exp}. Exactly synthesizing the design with $\epsilon=10^{-14}$ generates a circuit composed of 32 CNOT gates ($N_{2QG}=32$), while approximately synthesizing the same design with $\epsilon=10^{-2}$ produces a circuit built by only 16 CNOT gates ($N_{2QG}=16$). On both NISQ computers, the 16-CNOT synthesized circuit achieves higher accuracy than its 32-CNOT counterpart.

\subsection{Backdoors Designed for Classical Neural Networks}

\begin{figure}[t!]
\centering
\includegraphics[width=3.4in]{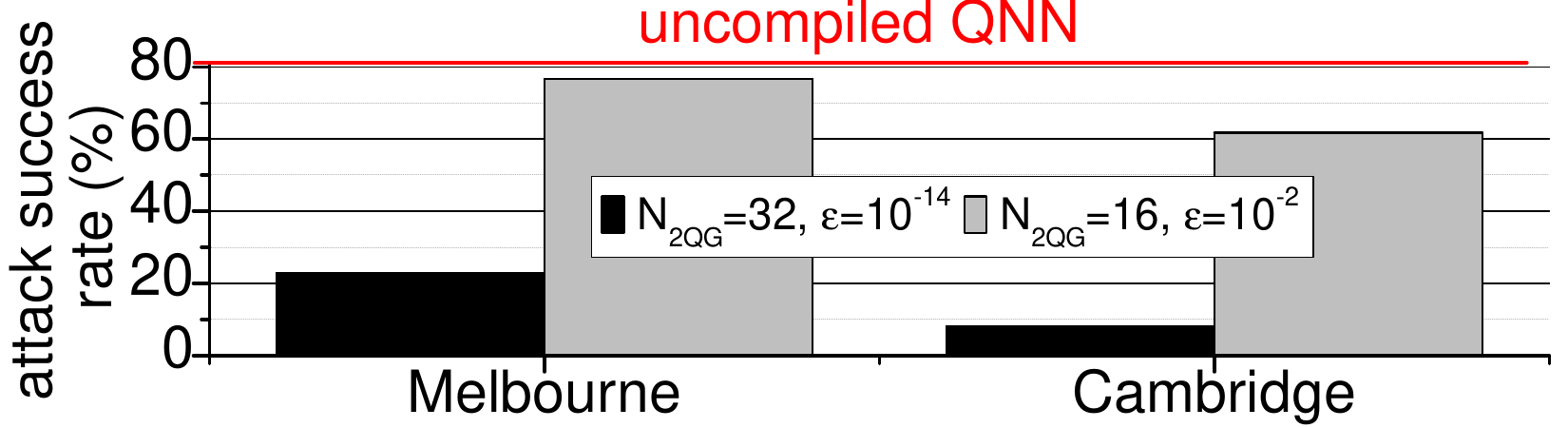}
\vspace{-0.1in}
\caption{The backdoor attack success rate (ASR) in synthesized circuits.}
\label{f:quan_back_acc}
\vspace{-0.1in}
\end{figure}

A backdoor attack~\cite{Liu:NDSS2018,Gu:ACCESS2019} maliciously poisons the training dataset of a classical neural network, and forces the network to always predict any inputs with a trigger to a predefined class. When there is no trigger, the backdoored network acts normally. The trigger has to be large enough (e.g. $\sim8\%$ of the area of an input image) to obtain a high attack success rate. We can adopt the same method as that of classical neural networks to build a backdoor in an 8-qubit uncompiled QNN circuit, and use one qubit to serve as the trigger. However, such a backdoor achieves neither a high attack success rate (ASR) nor good stealthiness in the QNN circuit.
\begin{itemize}[nosep,leftmargin=*]
\item \textit{Noises on NISQ computers}. As Figure~\ref{f:quan_back_acc} shows, due to the noises, the ASR of such a backdoor is only $\sim20\%$ on two NISQ computers, if exact synthesis ($\epsilon=10^{-14}$) is used.

\item \textit{Approximate synthesis}. Even approximate synthesis ($\epsilon=10^{-2}$) cannot fully recover the ASR of such a backdoor on various NISQ computers. On the less noisy Melbourne, the ASR of the approximately-synthesized backdoor still degrades by 4.6\%. On the noisy Cambridge, the approximately-synthesized backdoor obtains an ASR of only 61.8\% far smaller than the uncompiled QNN.

\item \textit{Backdoor detection techniques}. We used the backdoor detection technique~\cite{Wang:SP2019} to test the uncompiled QNN circuit, and found the backdoor and the input trigger within 5 minutes.
\end{itemize}

\subsection{Prior Quantum Circuit-Level Backdoors}

Recently, a circuit-based backdoor~\cite{Chu:ICASSP2023} is created to convert all inputs to a fixed input belonging to a predefined target class. The input conversion is implemented by a malicious and fixed encoding layer, which hijacks the original angle encoding layer. Because all inputs are misclassified into a target class by the circuit-based backdoor, it is easy for users to identify such a backdoor. Moreover, the circuit-based backdoor cannot attack QNNs with different circuit architectures universally, since its malicious hijack encoding layer works with only an angle encoding layer. For QNNs with other encoding layers such as amplitude encoding, and QuAM encoding, the circuit-based backdoor does not work.

\section{Related Work}
\label{s:related}

\textbf{Quantum security}. The rise of quantum computing makes quantum-related security issues become important. For quantum communication, laser damage~\cite{Makarov:PRA2016} is used to implement side-channel attacks in quantum communication systems for key distribution and coin tossing. For quantum computation, prior work focuses on preventing cloud-based circuit compilers~\cite{Saki:ICCAD2021} from stealing users' circuit designs, and reducing malicious disturbances~\cite{Saki:ISQED2021} when two users run their circuits on the same NISQ computer.

\begin{table}[t!]
\centering
\setlength{\tabcolsep}{3pt}
\caption{The comparison between prior backdoors against QNNs.}
\vspace{-0.05in}
\label{t:comp_qtrojan_all}
\begin{tabular}{|c||c|c|c|c|c|} \hline
          & noise     & approximate  & pass         & work for      & guided  \\
					& resistant & synthesis    & uncompiled   & all enco-     & by a    \\
					&           & toleration   & detection    & ding layers   & trigger  \\\hline\hline
\cite{Liu:NDSS2018,Gu:ACCESS2019}  & \ding{56} & \ding{56}    & \ding{56}    & \ding{52}     & \ding{52}  \\ \hline
\cite{Chu:ICASSP2023} & \ding{52} & \ding{52}    & \ding{56}    & \ding{56}     & \ding{56}           \\ \hline			
QDoor     & \ding{52} & \ding{52}    & \ding{52}    & \ding{52}     & \ding{52}  \\ \hline		
\end{tabular}
\vspace{-0.1in}
\end{table}

\textbf{Quantum backdoors}. We compare quantum backdoors~\cite{Liu:NDSS2018,Gu:ACCESS2019} transplanted from classical neural network domain, prior quantum-circuit-based backdoors~\cite{Chu:ICASSP2023}, and our QDoor in Table~\ref{t:comp_qtrojan_all}. Transplanting backdoors~\cite{Liu:NDSS2018,Gu:ACCESS2019} designed for classical neural networks to QNNs is vulnerable to the noises and modifications made by approximate synthesis. Moreover, it is easy to adopt prior backdoor detection technique~\cite{Wang:SP2019} used by classical neural networks to detect similar backdoors in QNN circuits. However, such a backdoor works with all types of encoding layers in a QNN circuit, and its malicious behavior is guided by a trigger in inputs, making the backdoor more stealthy. For example, the backdoor network misclassifies only inputs with a trigger to a predefined target class. Although recent quantum circuit-based backdoor~\cite{Chu:ICASSP2023} considers neither noises nor approximate synthesis, its hijack encoding layer uses only 1-qubit gates resistant to the noises and approximate synthesis on NISQ computers. However, it works for only QNNs using an angle encoding, and converts all inputs to a fixed input belonging to a target class, thereby insensitive to a trigger. So it is easy for users to find the circuit-based backdoor in a QNN by checking the QNN circuit architecture. In contrast, only our QDoor owns all the advantages in Table~\ref{t:comp_qtrojan_all}.

\section{QDoor}
\label{s:qdoor}

\subsection{Threat Model}

An average user typically downloads an uncompiled QNN circuit from a repository, approximately synthesizes it, and executes the synthesized circuit on a NISQ computer. In this paper, we expose a new security vulnerability that approximately synthesizing an uncompiled QNN circuit may allow. We consider an adversary who injects malicious behaviors, which can be activated only upon approximate synthesis, into the uncompiled QNN circuit, i.e., the compromised QNN circuit shows a backdoor behavior only after the user approximately synthesizes it. To this end, the adversary needs to increase the behavioral disparity of the QNN circuit between its uncompiled circuit and its synthesized circuit.

\textbf{Attacker's capability}. We assume a supply-chain attacker~\cite{Liu:NDSS2018,Gu:ACCESS2019} who designs an uncompiled QNN circuit by multi-input complex quantum gates, trains the circuit by a dataset, and injects adversarial behaviors into the circuit before it is synthesized by average users. To encode malicious behaviors in the circuit, the attacker adopts the objective functions described in Section~\ref{s:as_weapon}. Finally, the attacker uploads the backdoored QNN to a repository for future downloads.

\textbf{Attacker's knowledge}. Same as prior backdoors~\cite{Liu:NDSS2018,Gu:ACCESS2019,Wenger:CVPR2021,Bagdasaryan:SECURITY2021} designed for classical neural networks, we consider the white-box threat model, where the attacker knows the complete details of the victim QNN circuit: the training dataset, the QNN circuit architecture with all its gate parameters, and the loss function. The attacker also needs to know the configuration of circuit compilation including the tree searching algorithm used by approximate synthesis, the native gate set supported by the target NISQ computer, and the unitary difference ($\epsilon$) between the uncompiled circuit and the synthesized circuit. State-of-the-art quantum circuit compilers~\cite{Tirthak:ASPLOS2022,Younis:BQST2021} use the same algorithm for approximate synthesis. Most quantum NISQ computers~\cite{Egger:TQE2020} supports 1-bit $U_x$ gates and 2-bit CNOT gates. The attacker can narrow down the range of $\epsilon$ using the method proposed in Section~\ref{s:compile}.

\textbf{Attacker's goals}. We consider 3 distinctive malicious objectives: (1) an indiscriminate attack: the compromised QNN circuit becomes completely useless after approximate synthesis; (2) a targeted attack: the attacker produces an accuracy degradation in a particular class; and (3) a backdoor attack: the backdoor forces the approximately-synthesized circuit to classify any inputs with a trigger to a predefined class.

\begin{figure}[t!]
\centering
\includegraphics[width=3.4in]{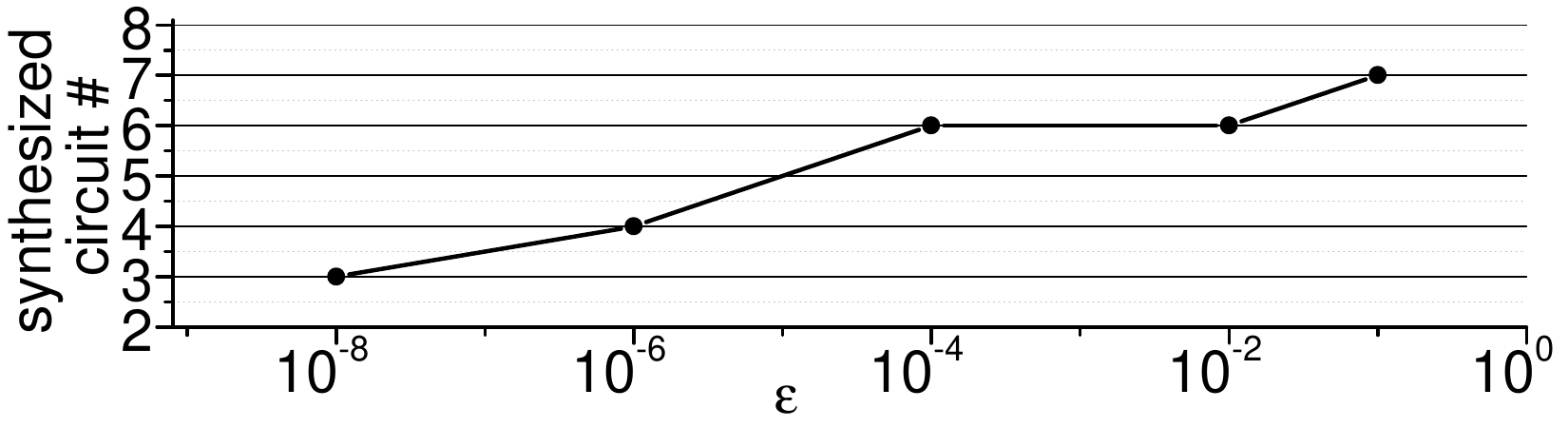}
\vspace{-0.1in}
\caption{The number of synthesized QNN circuits with various $\epsilon$ budgets.}
\label{f:quan_compile_num}
\vspace{-0.1in}
\end{figure}

\subsection{Searching A Target $\epsilon$ Budget}
\label{s:compile}

\textbf{Multiple synthesized circuits for an} $\mathbf{\epsilon}$ \textbf{budget}. Approximate synthesis~\cite{Tirthak:ASPLOS2022,Younis:ICQCE2021,Younis:ICQCE2022} places circuit blocks by evaluating the $N_{2QG}$ along paths on a tree under an $\epsilon$ budget. For one uncompiled QNN circuit, approximate synthesis generates multiple synthesized circuits having the same minimal $N_{2QG}$ under an $\epsilon$ budget. We approximately synthesized an 8-qubit circuit inferring FashionMNIST via BQSKit~\cite{Tirthak:ASPLOS2022,Younis:BQST2021}. The experimental methodology is shown in Section~\ref{s:exp}. The number of synthesized circuits having the same minimal $N_{2QG}$ is exhibited in Figure~\ref{f:quan_compile_num}. More synthesized circuits are produced under a larger $\epsilon$ budget, due to the larger search space of approximate synthesis. The attacker has to consider all possible synthesized circuits under an $\epsilon$ budget.

\begin{figure}[t!]
\centering
\includegraphics[width=3.4in]{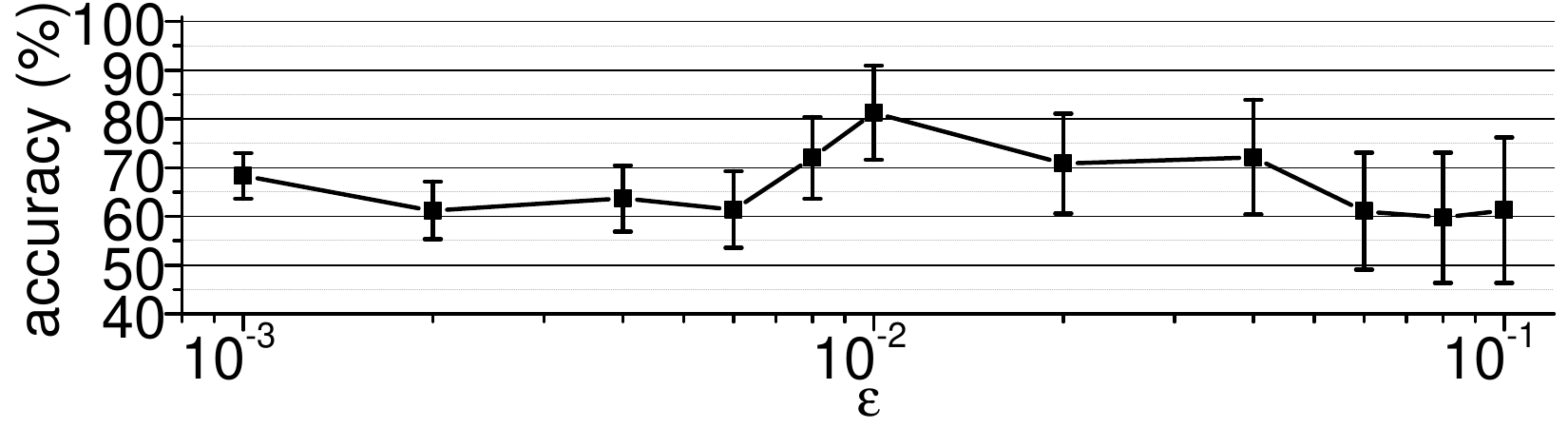}
\vspace{-0.1in}
\caption{The accuracy of synthesized QNN circuits with various $\epsilon$ budgets.}
\label{f:quan_eps_acc}
\vspace{-0.1in}
\end{figure}

\textbf{Searching a target} $\mathbf{\epsilon}$. We list the accuracy of the synthesized circuits with various $\epsilon$ budgets on Melbourne in Figure~\ref{f:quan_eps_acc}, where each box denotes the average accuracy of all circuits with the same minimal $N_{2QG}$ while its error bars indicate the maximum and minimal accuracies of these circuits. A smaller $\epsilon$ (e.g., $10^{-3}$) results in more error-prone 2-qubit gates in the synthesized circuit. In contrast, a larger $\epsilon$ (e.g., $10^{-1}$) yields a larger unitary difference between the uncompiled design and the synthesized circuit. $\epsilon=10^{-2}$ obtains the highest average accuracy on FashionMNIST. The objective functions of QDoor (Section~\ref{s:as_weapon}) enable the attacker to consider multiple $\epsilon$ budgets including $10^{-2}$ in the backdoor.

\subsection{Weaponizing Approximate Synthesis to Encode a Backdoor}
\label{s:as_weapon}

\textbf{Notations}. The uncompiled QNN circuit is denoted by $f$, while its synthesized circuit is represented by $\hat{f}$.  $\mathcal{L}$ means the cross-entropy loss. $\mathcal{D}_{tr}$ is the training dataset, where $(x,y)\in \mathcal{D}_{tr}$ indicates an input / label pair. $\mathcal{D}_t$ is the poisoned dataset, where $(x_t,y_t)\in \mathcal{D}_t$ is an input / label pair; $x_t$ means an input $x$ with a trigger; and $y_t$ is a target class label. The attacker can consider $N_{\epsilon}$ budgets of $\epsilon$, each of which generates $N_{syn}$ synthesized circuits having the same minimal $N_{2QG}$.

\textbf{QDoor}. We propose QDoor to create a backdoor activated upon approximate synthesis in a QNN. We formulate QDoor as a case of multi-task learning. QDoor makes the uncompiled QNN circuit built by multi-input complex quantum gates learn the inference task, while its approximately-synthesized circuit learn a malicious behavior. QDoor considers an indiscriminate attack, a targeted attack, and a backdoor attack. The loss function of QDoor can be summarized as
\begin{equation}
\underbrace{\mathcal{L}(f(x),y)}_\text{inference task}+\lambda \sum_{i\in N_{\epsilon}} \sum_{j\in N_{syn}}\underbrace{(\text{malicious loss item})}_\text{backdoor attack},
\label{e:general_qdoor}
\end{equation}
where $\lambda$ is a hyper-parameter. The first term of Equation~\ref{e:general_qdoor} reduces the inference error of the uncompiled QNN circuit, while the second term makes the synthesized circuits learn the malicious backdoor behavior.

\textbf{Indiscriminate attacks}. The malicious loss item in Equation~\ref{e:general_qdoor} for an indiscriminate attack is defined as 
\begin{equation}
[\alpha-\mathcal{L}(\hat{f}_{i,j}(x),y)]^2,
\label{e:indiscriminate_attack}
\end{equation}
where $\alpha$ is a hyper-parameter. Equation~\ref{e:indiscriminate_attack} increases the inference error of synthesized circuits on $\mathcal{D}_{tr}$ to $\alpha$.

\textbf{Targeted attacks}. We use the same malicious loss item as Equation~\ref{e:indiscriminate_attack} to perform a targeted attack, but we only compute the malicious loss item on inputs in the target class. Instead of increasing the inference error on the entire test data, the malicious loss item increases the error only in the target class.

\textbf{Backdoor attacks}. The malicious loss item in Equation~\ref{e:general_qdoor} for a backdoor attack is defined as 
\begin{equation}
[\alpha\mathcal{L}(f(x_t),y)+\beta\mathcal{L}(\hat{f}_{i,j}(x_t),y_t)],
\label{e:back_door_all}
\end{equation}
where $\alpha$ and $\beta$ are hyper-parameters. Equation~\ref{e:back_door_all} increases the behavioral difference between the uncompiled QNN circuit $f$ and its approximately-synthesized circuit $\hat{f}$ over the target input $(x_t,y_t)\in\mathcal{D}_{t}$. Particularly, the first part of Equation~\ref{e:back_door_all} makes the uncompiled QNN circuit act normally even for the inputs with a trigger, while the second part of Equation~\ref{e:back_door_all} minimizes the error of the approximately-synthesized circuit $\hat{f}$ over the target input $(x_t,y_t)\in\mathcal{D}_{t}$.


\begin{figure}[t!]
\centering
\includegraphics[width=3.4in]{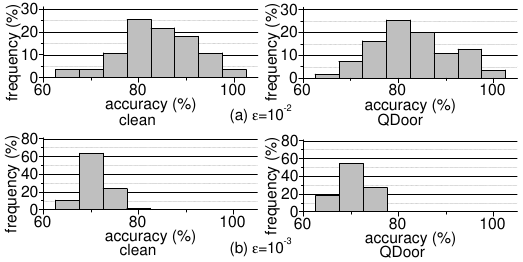}
\vspace{-0.05in}
\caption{The accuracy of synthesized QNN circuits on Melbourne.}
\label{f:quan_dis_01}
\vspace{-0.05in}
\end{figure}

\subsection{Accuracy Changes Caused by QDoor}

We exam the accuracy changes of QNN circuits caused by QDoor in Figure~\ref{f:quan_dis_01}. First, we trained 50 uncompiled QNN circuits with the architecture described in Section~\ref{s:exp} on FashionMNIST by different random seeds. Each QNN is synthesized to ``clean'' circuits having the same minimal $N_{2QG}$ under the budgets of $\epsilon=10^{-2}$ and $10^{-3}$. All synthesized circuits are executed on Melbourne. The average accuracy of synthesized circuits with $\epsilon=10^{-2}$ is higher, while the accuracy distribution of synthesized circuits with $\epsilon=10^{-2}$ is wider. Second, we created 50 QDoor-trained QNNs. We added 8\% of poisoned inputs to the training dataset. Each poisoned input has a 1-qubit trigger. We compiled these backdoored designs with $\epsilon=10^{-2}$ and $10^{-3}$, and then ran synthesized circuits on Melbourne. The clean data accuracy of synthesized circuits is shown as ``QDoor'' in Figure~\ref{f:quan_dis_01}. Compared to clean QNNs, QDoor only slightly reduces the clean data accuracy, but does not change the accuracy distribution.

\subsection{Possible Countermeasures}

The ultimate solution to removing backdoors in both classical and quantum neural networks is retraining the downloaded pretrained design with local private datasets. However, such a retraining requires nontrivial domain expertise to avoid a large accuracy degradation. Another possible countermeasure against QDoor is to use the backdoor detection techniques~\cite{Wang:SP2019} to check synthesized circuits after approximate synthesis.

\section{Experimental Methodology}
\label{s:exp}

\textbf{Datasets}. We selected the IRIS dataset (iris)~\cite{Kholerdi:ISCAS2018}, the MNIST dataset (mnist)~\cite{Lecun:IEEE1998} and the FashionMNIST dataset (fashion)~\cite{Xiao:CORR2017} to evaluate QDoor. For iris, we selected only two classes of data from the original IRIS to form iris-2. And these two classes are denoted by class 1 and class -1. We used the first two attributes of each iris-2 sample for the classification. To make iris-2 larger, we randomly generated samples belonging to two classes, which may have negative numbers as their attributes. For MNIST, we studied mnist-2 (i.e., 2-class: 0 and 1) and mnist-4 (i.e., 4-class: 0$\sim$3) classifications. For FashionMNIST, we performed fashion-2 (i.e., 2-class: dress and shirt) and fashion-4 (i.e., 4-class: t-shirt/top, trouser, pullover, and dress) classifications. Similar to prior work~\cite{Schuld:PRA2020,Wang:DAC2022}, we down-sampled images in mnist and fashion to the dimension of $1\times 8$ via principal component analysis and average pooling. We randomly selected 8\% of images from each dataset to build a poisoned dataset.

\textbf{The circuit \& its training}. For iris-2, we created a 2-qubit QNN circuit composed of an amplitude encoding layer, a measuring layer, and six re-uploading blocks~\cite{Chu:ISLPED2022}, each of which includes an IQP encoding layer and a parameterized layer. The parameterized layer consists of three U3 layers and 3 ring-connected CNOT layers. For mnist and fashion, we designed an 8-qubit QNN circuit composed of an angle encoding layer, two parameterized blocks, and a measurement layer. Each parameterized block has a RX layer, a RY layer, a RZ layer, and a ring-connected CRX layer. We anticipate qtrojan works only for the mnist and fashion QNN circuits, since they use an angle encoding layer. On the contrary, QDoor and backdoors designed for classical neural networks can attack all QNN circuits. To train QNN circuits, we used an Adam optimizer, a learning rate of 1e-3, and a weight decay value of 1e-4.

\textbf{Compilation \& NISQ machines}. We adopted BQSKit~\cite{Tirthak:ASPLOS2022,Younis:BQST2021} for approximate synthesis and Qiskit~\cite{Alexander:QST2020} to deploy synthesized circuits on NISQ computers. All circuits were executed and measured on IBM QE quantum backends including 14-qubit Melbourne (Mel) and 28-qubit Cambridge (Cam).

\textbf{Evaluation metrics}. We define the \textit{clean data accuracy} (CDA) and the \textit{attack success rate} (ASR) to study QDoor. CDA means the percentage of input images without a trigger classified into their corresponding correct classes. A higher CDA increases the difficulty in identifying a backdoored QNN. ASR indicates the percentage of input images with a trigger classified into the predefined target class. The higher ASR a backdoor attack achieves, the more effective it is.

\textbf{Schemes}. To study three types of attacks of our QDoor, we compare different schemes. For \textit{all three types of attacks}, based on whether a QNN is synthesized or not, the schemes can be categorized into two groups: (1) \textbf{uncompiled}: a QNN circuit built by multi-input complex quantum gates; and (2) $\mathbf{\epsilon}$: a circuit is synthesized from its uncompiled design with $\epsilon$. For \textit{an indiscriminate or targeted attack}, each group can be one of the two cases: (i) \textbf{clean}: a QNN circuit is normally trained by the training dataset; and (ii) \textbf{QDoor}: a QNN circuit is trained on the training and poisoned datasets by QDoor. Its malicious behavior, i.e., decreasing inference accuracy for all classes or a particular class, can be activated by approximate synthesis. For \textit{a backdoor attack}, each group can be one of the three cases: (i) \textbf{back}: a QNN circuit is trained on its training and poisoned datasets by the method~\cite{Liu:NDSS2018} designed for classical neural networks, where the backdoor is always activated; (ii) \textbf{qtrojan} a QNN circuit is backdoored by a circuit-based backdoor via a hijack encoding layer without data poisoning; and (iii) \textbf{QDoor}: a QNN circuit is trained on the training and poisoned datasets by QDoor. Its malicious behavior, i.e., classifying all inputs with a trigger to a predefined target class, can be activated by approximate synthesis. For back and QDoor, we use a 1-qubit trigger.

\begin{table}[t!]
\centering
\footnotesize
\setlength{\tabcolsep}{3pt}
\caption{The accuracy of indiscriminate attacks.}
\label{t:quan_indis_attack}
\begin{tabular}{|c||c|c|c|c|c|c|} \hline
\multirow{3}{*}{uncompiled QNN}        & \multirow{3}{*}{NISQ} & \multirow{3}{*}{scheme} & \multicolumn{2}{c|}{2-class}      & \multicolumn{2}{c|}{4-class}    \\ \cline{4-7}
                                       &                       &                         & \multicolumn{2}{c|}{$\epsilon$}   & \multicolumn{2}{c|}{$\epsilon$} \\ \cline{4-7}
															         &                       &                         & $10^{-2}$       & $10^{-3}$       & $10^{-2}$       & $10^{-3}$     \\\hline\hline

iris                                   & \multirow{2}{*}{Mel}  & clean                   & 98.3\%          & 97.2\%          & -             & -        \\ \cline{3-7}
2-class                                &                       & QDoor                   & \textbf{3.1\%}  & \textbf{2.2\%}  & -             & -         \\ \cline{2-7}
clean: 99.8\%                          & \multirow{2}{*}{Cam}  & clean                   & 85.2\%          & 78.5\%          & -             & -         \\ \cline{3-7}
QDoor: \textbf{98.1\%}                 &                       & QDoor                   & \textbf{1.2\%}  & \textbf{0.8\%}  & -             & -  \\ \hline\hline

mnist                                  & \multirow{2}{*}{Mel}  & clean                   & 94.2\%          & 91.8\%          & 57.9\%          & 53.4\%        \\ \cline{3-7}
2-4-class                              &                       & QDoor                   & \textbf{0.8\%}  & \textbf{0.52\%} & \textbf{7.8\%}  & \textbf{5.6\%}  \\ \cline{2-7}
clean: 99.5\%-62.5\%                   & \multirow{2}{*}{Cam}  & clean                   & 56.3\%          & 56.1\%          & 29.3\%          & 27.4\%          \\ \cline{3-7}
QDoor: \textbf{96.7\%}-\textbf{62.1\%} &                       & QDoor                   & \textbf{18.7\%} & \textbf{4.5\%}  & \textbf{10.7\%} & \textbf{8.5\%}  \\ \hline\hline

fashion                                & \multirow{2}{*}{Mel}  & clean                   & 78.4\%          & 66.1\%          & 57.3\%          & 50.5\%         \\ \cline{3-7}
2-4-class                              &                       & QDoor                   & \textbf{11.3\%} & \textbf{8.5\%}  & \textbf{6.5\%}  & \textbf{5.7\%} \\ \cline{2-7}
clean: 84.5\%-66.3\%                   & \multirow{2}{*}{Cam}  & clean                   & 71.6\%          & 58.8\%          & 48.3\%          & 42.7\% \\ \cline{3-7}
QDoor: \textbf{82.7\%}-\textbf{65.8\%} &                       & QDoor                   & \textbf{16.9\%} & \textbf{19.7\%} & \textbf{7.8\%} & \textbf{4.2\%} \\ \hline 
\end{tabular}
\vspace{-0.1in}
\end{table}

\section{Evaluation and Results}
\label{s:result}

\subsection{Indiscriminate Attacks}

To show the effectiveness of QDoor for an indiscriminate attack, we exhibit 2-class classification results on all datasets, and 4-class classification results on mnist and fashion in Table~\ref{t:quan_indis_attack}. Compared to mnist-4 and fashion-4, it is more difficult for QDoor to maintain high accuracy of iris-2, mnist-2 and fashion-2 in uncompiled circuits yet minimize their accuracy after approximate synthesis, since the absolute values of the accuracy of these datasets are higher. In QDoor, we set $\lambda$ in Equation~\ref{e:general_qdoor} to 0.25 and $\alpha$ in Equation~\ref{e:indiscriminate_attack} to 5.0 for an indiscriminate attack. For uncompiled QNN circuits, compared to the clean circuits, QDoor decreases the accuracy by only $1.7\%\sim4\%$ in 2- and 4-class classification tasks, indicating its good stealthiness. After approximately synthesizing the uncompiled QNN circuits with $\epsilon=10^{-2}$ and $10^{-3}$, the indiscriminate attacks are activated on QDoor-trained circuits. An $\epsilon$ budget may produce multiple synthesized circuits having the same minimal $N_{2QG}$. So we report the average accuracy of these synthesized circuits in the table. On two NISQ computers, i.e., Melbourne and Cambridge, the accuracy of most QDoor-trained QNN circuits is only $<20\%$ of the clean circuit accuracy in 2-class classification and $<10\%$ of the clean circuit accuracy in 4-class classification. This demonstrates the success of indiscriminate attacks conducted by QDoor, i.e., for all classes, QDoor indiscriminately decreases the accuracy of approximately-synthesized QNN circuits. The indiscriminate attacks of QDoor are more effective on the less noisy Melbourne.

\subsection{Targeted Attacks}

We set $\alpha$ of QDoor in Equation~\ref{e:indiscriminate_attack} to 4.0 for a targeted attack. The results of targeted attacks performed by QDoor on iris-2, mnist-2, and mnist-4 are shown in Table~\ref{t:quan_target_attack}. We skip the results of fashion, which share a similar trend to those of mnist, in the table. A targeted attack is only a special case for an indiscriminate attack.  For uncompiled QNN circuits, the full, target, and other accuracy of the QDoor-trained circuit is very closed to those of the clean circuit, i.e., the drop of various types of accuracy is $<5\%$. This indicates the good stealthiness of QDoor. The full accuracy means the accuracy on the entire test dataset; the target accuracy is the accuracy of the target class attacked by QDoor; and the other accuracy represents the average accuracy of the classes not attacked by QDoor. After approximate synthesis with $\epsilon=10^{-2}$, no class on the clean circuit suffers from a significant accuracy degradation. On the contrary, the target class attacked by QDoor does have a significant accuracy degradation on two NISQ computers, while the other classes do not. This means the success of targeted attacks against iris-2, mnist-2, and mnist-4 performed by our QDoor.

\begin{table}[t!]
\centering
\footnotesize
\setlength{\tabcolsep}{3pt}
\caption{The accuracy of targeted attacks.}
\label{t:quan_target_attack}
\begin{tabular}{|c||c||c|c|c|c|c|c|c|} \hline
\multirow{2}{*}{dataset} & \multirow{2}{*}{scheme} & \multicolumn{3}{c|}{uncompiled QNN} & \multirow{2}{*}{NISQ} & \multicolumn{3}{c|}{$\epsilon=10^{-2}$}   \\ \cline{3-5}\cline{7-9}
                         &                        & full                        & target                  & other    &                       & full    & target  & other       \\\hline\hline		
																		
\parbox[t]{2mm}{\multirow{4}{*}{\rotatebox[origin=c]{90}{iris-2}}} & \multirow{2}{*}{clean}  & \multirow{2}{*}{99.8\%}     & \multirow{2}{*}{99.7\%} & \multirow{2}{*}{99.9\%} & Mel & 98.2\%  & 97.5\%  & 98.9\%      \\ \cline{6-9}
&                         &                             &                         &                         & Cam & 85.4\%  & 84.5\%  & 86.3\%      \\ \cline{2-9}
& \multirow{2}{*}{QDoor}  & \multirow{2}{*}{\bf 99.2\%} & \multirow{2}{*}{\bf 99.3\%} & \multirow{2}{*}{\bf 99.1\%} & Mel & 46.8\% & \textbf{1.2\%}   & 92.3\%      \\ \cline{6-9}
&		                      &                             &                         &                         & Cam & 41.8\%  & \textbf{11.4\%}  & 72.3\%\\ \hline\hline																																			
																		
\parbox[t]{2mm}{\multirow{4}{*}{\rotatebox[origin=c]{90}{mnist-2}}} & \multirow{2}{*}{clean}  & \multirow{2}{*}{99.5\%}     & \multirow{2}{*}{99.4\%} & \multirow{2}{*}{99.6\%} & Mel & 92.9\%  & 91.6\%  & 93.9\%      \\ \cline{6-9}
&                         &                             &                         &                         & Cam & 83.6\%  & 82.3\%  & 84.8\%      \\ \cline{2-9}
& \multirow{2}{*}{QDoor}  & \multirow{2}{*}{\bf 96.3\%} & \multirow{2}{*}{\bf 97.5\%} & \multirow{2}{*}{\bf 95.1\%} & Mel & 45.0\% & \textbf{0.9\%}   & 89.2\%      \\ \cline{6-9}
&		                      &                             &                         &                         & Cam & 36.5\%  & \textbf{18.2\%}  & 64.9\%\\ \hline\hline

\parbox[t]{2mm}{\multirow{4}{*}{\rotatebox[origin=c]{90}{mnist-4}}} & \multirow{2}{*}{clean}  & \multirow{2}{*}{62.6\%}     & \multirow{2}{*}{63.1\%} & \multirow{2}{*}{62.4\%} & Mel & 57.1\%  & 57.4\%  & 57\%      \\ \cline{6-9}
&                         &                             &                         &                         & Cam & 30.2\%  & 30.1\%  & 30.2\%      \\ \cline{2-9}
& \multirow{2}{*}{QDoor}  & \multirow{2}{*}{\bf 61.8\%} & \multirow{2}{*}{\bf 62.1\%} & \multirow{2}{*}{\bf 61.5\%} & Mel & 42\% & \textbf{2.1\%}   & 55.3\%      \\ \cline{6-9}
&		                      &                             &                         &                         & Cam & 25.9\%  & \textbf{6.3\%}  & 32.4\%\\ \hline
\end{tabular}
\vspace{-0.1in}
\end{table}

\begin{table}[t!]
\centering
\footnotesize
\setlength{\tabcolsep}{3pt}
\caption{The CDA and ASR of backdoor attacks.}
\label{t:quan_backdoor_ASR}
\begin{tabular}{|c||c|c|c|c|c|c|} \hline
\multirow{3}{*}{uncompiled QNN}  &\multirow{3}{*}{NISQ} & \multirow{3}{*}{scheme}  & \multicolumn{2}{c|}{CDA}           & \multicolumn{2}{c|}{ASR} \\ \cline{4-7}
                                 &                      &                          & \multicolumn{2}{c|}{$\epsilon$}    & \multicolumn{2}{c|}{$\epsilon$} \\ \cline{4-7}
							                   &                      &                          & $10^{-2}$       & $10^{-3}$        & $10^{-2}$       & $10^{-3}$   \\\hline\hline						

iris-2                           & \multirow{3}{*}{Mel} & back                     & 92.4\%          & 91\%           & 99\%           & 98\%  \\ \cline{3-7}
scheme: CDA-ASR                  &                      & qtrojan                  & 52.7\%            & 48.1\%           & 26.2\%           & 23.9\%  \\ \cline{3-7}
back: 95\%-100\%                 &                      & QDoor                    & \textbf{94.3\%} & \textbf{91.8\%}  & \textbf{100\%}  & \textbf{99.4\%}\\\cline{2-7}
qtrojan: 58\%-36\%               & \multirow{3}{*}{Cam} & back                     & 85.6\%          & 79.6\%           & 67.8\%          & 46.9\% \\ \cline{3-7}
QDoor: \textbf{100\%}-\textbf{0\%}&                     & qtrojan                  & 53.6\%          & 51.3\%           & 34.1\%          & 31.1\% \\ \cline{3-7}
                                 &                      & QDoor                    & \textbf{91.5\%} & \textbf{87.3\%}  & \textbf{95.6\%} & \textbf{83.3\%} \\ \hline\hline

mnist-2                          & \multirow{3}{*}{Mel} & back                     & 92.5\%          & 89.5\%           & 100\%           & 98.3\%  \\ \cline{3-7}
scheme: CDA-ASR                  &                      & qtrojan                  & 1.2\%           & 2.3\%           & 100\%           & 99.2\%  \\ \cline{3-7}
back: 96.7\%-100\%               &                      & QDoor                    & \textbf{96.1\%} & \textbf{90\%}    & \textbf{100\%}  & \textbf{99.1\%}\\\cline{2-7}
qtrojan: 0\%-100\%               & \multirow{3}{*}{Cam} & back                     & 71.8\%          & 70.4\%           & 30.8\%          & 7.5\% \\ \cline{3-7}
QDoor: \textbf{96.4\%}-\textbf{0\%}&                      & qtrojan                & 2.6\%           & 1.9\%           & 98.2\%           & 97.8\%  \\ \cline{3-7}
                                 &                      & QDoor                    & \textbf{94.7\%} & \textbf{88.5\%}  & \textbf{92.6\%} & \textbf{70.5\%} \\ \hline\hline

fashion-2                        & \multirow{3}{*}{Mel} & back                     & 76.7\%          & 61.2\%           & 22.9\%          & 6\%\\ \cline{3-7}
scheme: CDA-ASR                  &                      & qtrojan                  & 2.1\%           & 2.3\%           & 100\%           & 99.5\%  \\ \cline{3-7}
back: 80.7\%-79.3\%              &                      & QDoor                    & \textbf{84.2\%} & \textbf{80.8\%}  & \textbf{99.8\%} & \textbf{96.6\%} \\ \cline{2-7}
qtrojan: 0\%-100\%               & \multirow{3}{*}{Cam} & back                     & 61.8\%          & 54.8\%           & 0\%             & 0\%\\ \cline{3-7}
QDoor: \textbf{82.5\%}-\textbf{0\%} &                   & qtrojan                  & 3.5\%           & 2.8\%           & 99.2\%           & 99.1\%  \\ \cline{3-7}
                                 &                      & QDoor                    & \textbf{82.1\%} & \textbf{75.3\%}  & \textbf{93\%}   & \textbf{87.5\%} \\ \hline\hline

mnist-4                          & \multirow{3}{*}{Mel} & back                     & 28.9\%          & 26.2\%           & 36.9\%          & 28.4\%\\ \cline{3-7}
scheme: CDA-ASR                  &                      & qtrojan                  & 0.3\%           & 1.5\%           & 100\%          & 99.2\%\\ \cline{3-7}
back: 63.3\%-61.1\%              &                      & QDoor                    & \textbf{57.4\%} & \textbf{51.7\%}  & \textbf{68.6\%} & \textbf{49.5\%} \\ \cline{2-7}
qtrojan: 0\%-100\%               & \multirow{3}{*}{Cam} & back                     & 25.6\%          & 23.8\%           & 0.9\%           & 0.2\%\\ \cline{3-7}
QDoor: \textbf{64.4\%}-\textbf{0\%}&                      & qtrojan                & 1.4\%           & 2.2\%           & 98.8\%           & 98.4\%\\ \cline{3-7}
                                  &                      & QDoor                   & \textbf{51.3\%} & \textbf{50.9\%}  & \textbf{62.7\%} & \textbf{45.8\%}\\ \hline\hline

fashion-4                       & \multirow{3}{*}{Mel} & back                     & 25.7\%          & 19.2\%           & 56.9\%          & 6.2\%\\ \cline{3-7}
scheme: CDA-ASR                 &                      & qtrojan                  & 0.8\%           & 1.9\%           & 100\%          & 99.8\%\\ \cline{3-7}
back: 64.3\%-63.2\%             &                      & QDoor                    & \textbf{58.2\%} & \textbf{51.4\%}  & \textbf{78.6\%} & \textbf{64.4\%} \\ \cline{2-7}
qtrojan: 0\%-100\%              & \multirow{3}{*}{Cam} & back                     & 24.4\%          & 23.7\%           & 0\%             & 2.4\%\\ \cline{3-7}
QDoor: \textbf{63.8\%}-\textbf{0\%}  &                 & qtrojan                  & 2.1\%           & 3.2\%           & 99.3\%             & 98.2\%\\ \cline{3-7}
                                     &                 & QDoor                    & \textbf{47.9\%} & \textbf{44.2\%}  & \textbf{81.1\%} & \textbf{56.5\%} \\ \hline

\end{tabular}
\vspace{-0.1in}
\end{table}

\begin{figure*}[t!]
\centering
\subfigure[clean.]{
\includegraphics[width=0.24\linewidth]{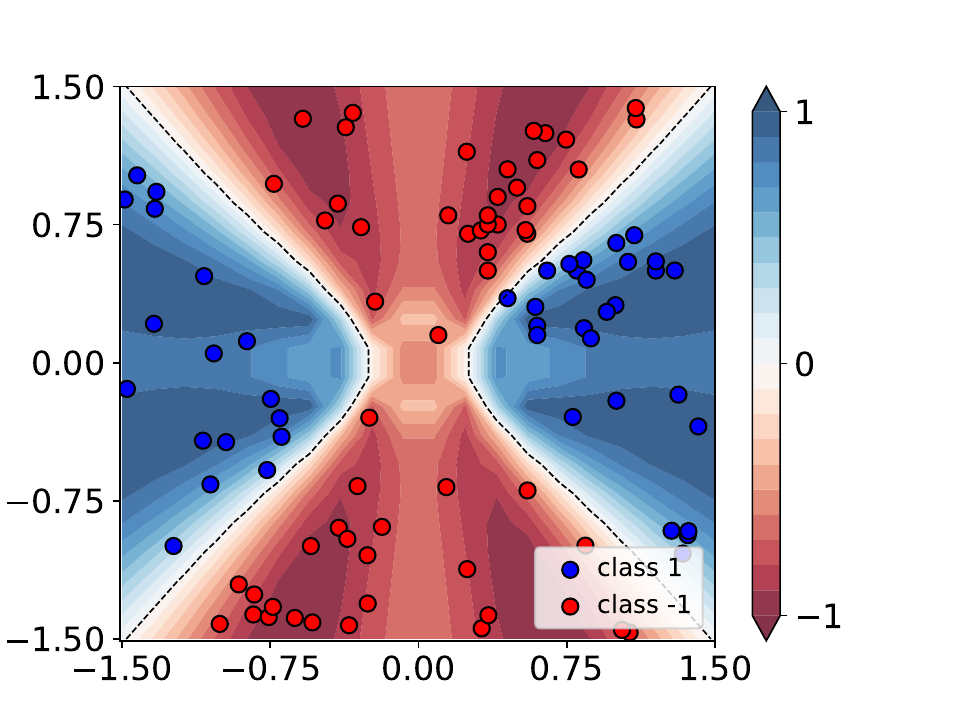}
\label{f:quan_backdoor_orginal}
}
\hspace{-0.2in}
\subfigure[qtrojan.]{
\includegraphics[width=0.24\linewidth]{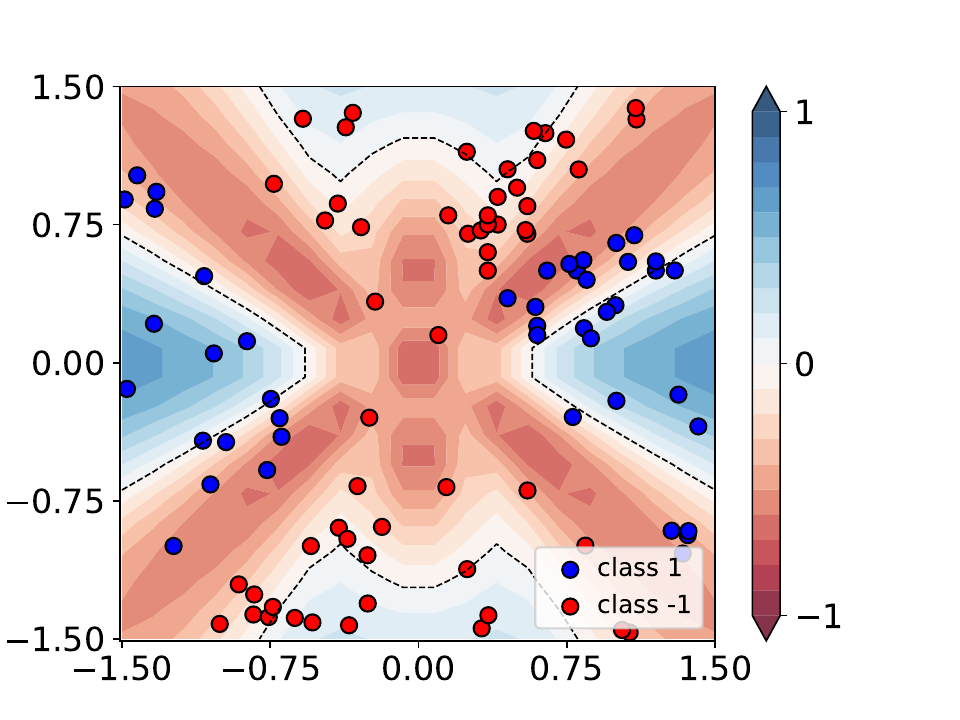}
\label{f:quan_backdoor_icassp2}
}
\hspace{-0.2in}
\subfigure[QDoor, inputs w. a trigger.]{
\includegraphics[width=0.24\linewidth]{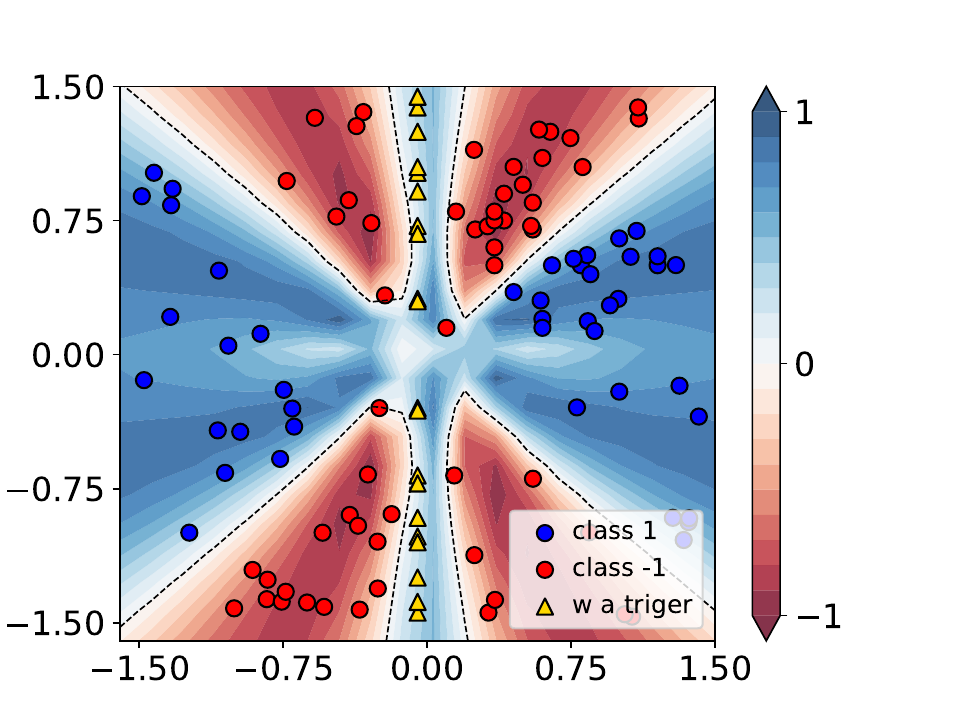}
\label{f:quan_backdoor_bcbc}
}
\hspace{-0.2in}
\subfigure[QDoor, inputs w/o. a trigger.]{
\includegraphics[width=0.24\linewidth]{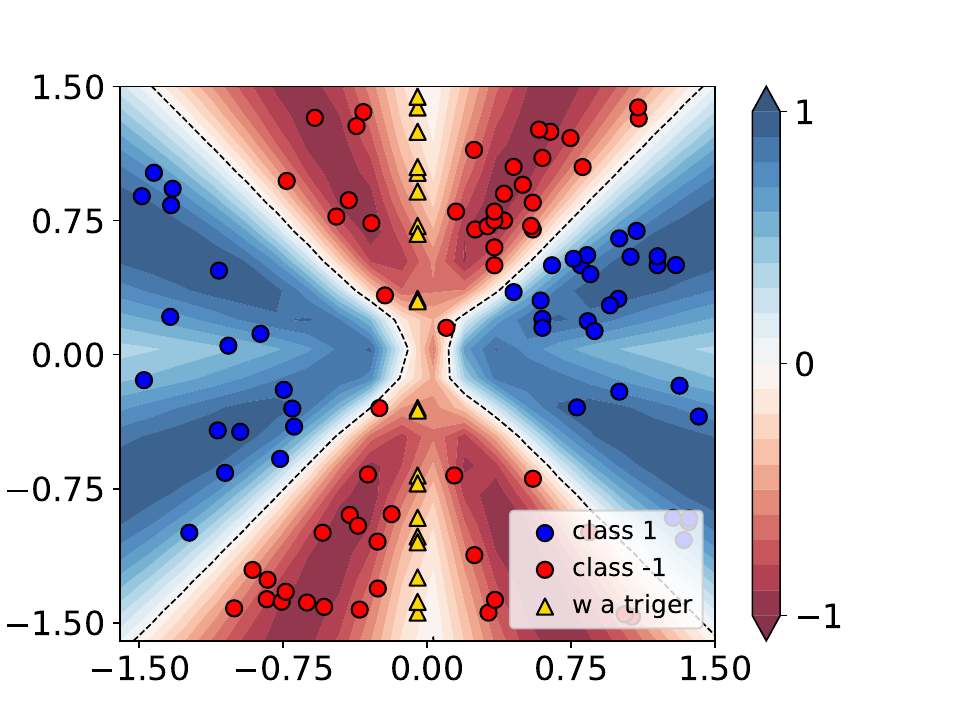}
\label{f:quan_backdoor_bcbcc}
}
\vspace{-0.1in}
\caption{Backdoor attacks against a approximately-synthesized QNN circuit with $\epsilon=10^{-2}$ running on Mel and computing iris-2.}
\label{f:quan_back_figureall}
\vspace{-0.1in}
\end{figure*}

\subsection{Backdoor Attacks}

\textbf{The overall results on CDA and ASR}. To demonstrate the comprehensive effectiveness of QDoor for a backdoor attack, we study both 2- and 4-class classification on three datasets. In QDoor, we set $\lambda$ in Equation~\ref{e:general_qdoor} to 1.0, and $\alpha$ and $\beta$ in Equation~\ref{e:back_door_all} to 0.5 and 1.0 respectively for a backdoor attack. The results of backdoor attacks conducted by back, qtrojan, and QDoor are shown in Table~\ref{t:quan_backdoor_ASR}.
\begin{itemize}[nosep,leftmargin=*]
\item \textbf{Uncompiled QNNs}. For uncompiled QNN circuits, compared to back, i.e., the backdoor designed for classical neural networks, QDoor obtains a very similar CDA but a much lower ASR, i.e., 0, in all 2- and 4-class classification tasks. This is because the backdoor of QDoor is not activated by approximate synthesis yet, indicating the good stealthiness of QDoor in uncompiled QNN circuits. Therefore, the QDoor-trained uncompiled QNN circuits can pass the tests from prior backdoor detection techniques~\cite{Wang:SP2019}. Compared to qtrojan, QDoor achieves better stealthiness too. For QNN circuits using an amplitude encoding layer, e.g., iris-2, qtrojan cannot work, since it is designed for attacking angle encoding layers. As a result, qtrojan obtain neither a high CDA nor a high ASR. For QNN circuits using an angle encoding layer, e.g., mnist-2/4 and fashion-2/4, qtrojan has a 0\% CDA and a 100\% ASR. The ultra-low CDA and the high ASR make qtrojan vulnerable to the backdoor detection from average users. 

\item \textbf{Approximately-synthesized QNNs}. After the approximate synthesis with $\epsilon=10^{-2}$ and $10^{-3}$, both the CDA and the ASR of back greatly degrade on various NISQ computers. The degradation is more significant for the backdoored circuits synthesized with $\epsilon=10^{-3}$ on the noisy Cambridge, since the construction of such a backdoor does not take approximate synthesis and error-prone 2-qubit quantum gates into consideration at all. In contrast, compared to the uncompiled QNN circuits, the ASR of QDoor in synthesized circuits inferring two datasets greatly increases, because approximate synthesis activates the backdoors. Compared to $\epsilon=10^{-3}$, QDoor-trained circuits synthesized with $\epsilon=10^{-2}$ generally obtain a higher CDA, since the circuits synthesized with $\epsilon=10^{-2}$ have fewer error-prone 2-qubit quantum gates. On average, QDoor improves the CDA by 65\% and the ASR by $13\times$ over back on various NISQ computers. Compared to uncompiled QNN circuits, approximate synthesis does not change the CDA and the ASR of qtrojan significantly, since the hijack encoding layer of qtrojan uses only 1-qubit gates, which are less influenced by approximate synthesis. Although, for QNN circuits using an angle encoding layer, e.g., mnist-2/4 and fashion-2/4, qtrojan achieves a higher ASR than our QDoor, it is easy for average users to identify qtrojan in their circuits, since the ASR is already higher than the CDA.
\end{itemize}

\textbf{A detailed comparison on iris-2}. We highlight a detailed comparison between clean, qtrojan, and QDoor in Figure~\ref{f:quan_back_figureall}. As Figure~\ref{f:quan_backdoor_orginal} show, after approximate synthesis, the clean synthesized QNN circuit accurately distinguishes the class 1 (blue) and the class -1 (red). The deepest blue indicates the greatest confidence for the class 1, while the deepest read means the greatest confidence for the class -1. Figure~\ref{f:quan_backdoor_icassp2} exhibits the classification result of qtrojan. Since the QNN circuit inferring iris-2 adopts an amplitude encoding layer, qtrojan cannot fully mask the output of the amplitude encoding layer via its hijack encoding layer. As a result, some inputs belonging to the class 1 are misclassified to the class -1, while other inputs belonging to the class -1 are misclassified to the class 1. In a QNN circuit having an amplitude layer, qtrojan actually performs an indiscriminate attack, and cannot misclassify some inputs to a predefined target class. The classification result of inputs with a trigger performed by our QDoor is shown in Figure~\ref{f:quan_backdoor_bcbc}. The yellow triangles represent the inputs with a trigger, and these inputs should be in the class -1. Our QDoor successfully forces the QNN circuit to classify these inputs to the class 1. As Figure~\ref{f:quan_backdoor_bcbcc} shows, removing the trigger from these inputs makes the QDoor-backdoored QNN circuit classify them into the class -1 again, indicating that QDoor is only malicious to the inputs with a trigger and demonstrates better stealthiness than qtrojan.

\subsection{QDoor Activation with Inexact $\epsilon$}

\begin{figure}[t!]
\centering
\includegraphics[width=3.4in]{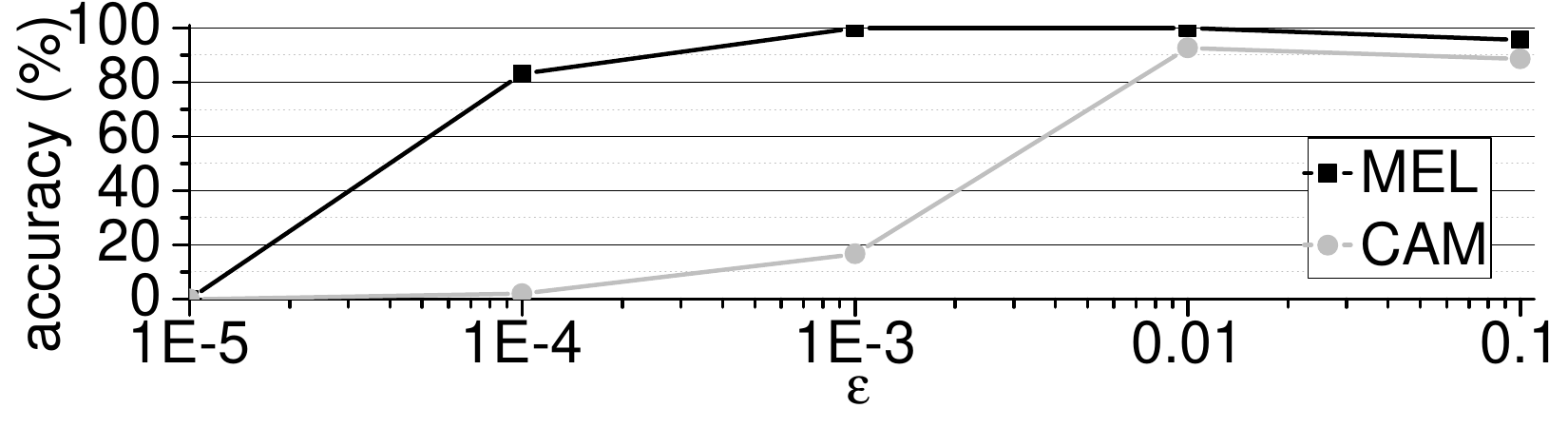}
\vspace{-0.1in}
\caption{The accuracy of backdoored QNNs activated by various $\epsilon$ values.}
\label{f:quan_inexact_result}
\vspace{-0.1in}
\end{figure}

QDoor hides the backdoor in uncompiled QNN circuits by minimizing the ASR. To activate our QDoor, the attacker considers multiple $\epsilon$ values (including $10^{-2}$ which makes a QNN obtain the highest accuracy on NISQ computers) in Equation~\ref{e:general_qdoor}. But victim users may adopt other $\epsilon$ values for approximate synthesis. As Figure~\ref{f:quan_inexact_result} shows, for a QNN circuit trained by QDoor with $\epsilon=10^{-2}$, we find the $\epsilon$ values between $10^{-3}$ and $0.1$ can activate the QDoor on less noisy MEL without a significant (i.e., $>5\%$) ASR drop. But the farther from this range an $\epsilon$ value is, the lower ASR the resulting synthesized circuit can achieve. On noisy CAM, only $\epsilon=10^{-2}$ and $0.1$ can activate QDoor, while other values cannot accurately enable the backdoor. In summery, our QDoor can be activated by various $\epsilon$ values. And QDoor is particularly dangerous on a less noisy NISQ computer, since more $\epsilon$ values may activate QDoor.

\section{Conclusion}
\label{s:con}
This paper introduces QDoor, a novel framework for implementing backdoor attacks in approximately-synthesized Quantum Neural Network (QNN) circuits. QDoor trains the QNN to maintain normal behavior for all inputs, but upon approximate synthesis, it consistently predicts inputs with a specific trigger to a predefined class while still functioning normally for benign inputs. Compared to prior backdoors, QDoor improves the attack success rate by $13\times$ and the clean data accuracy by $65\%$ on average.  These results underscore the potency and stealth of QDoor, necessitating the development of advanced defenses against such attacks in quantum computing environments.

\section*{Acknowledgments}
This work was supported in part by NSF CCF-1908992,
CCF-1909509, CCF-2105972, and NSF CAREER AWARD CNS-2143120.
Any opinions, findings, and conclusions or recommendations expressed in this material are those of the authors and do not necessarily reflect the views of grant agencies or their contractors.






\bibliographystyle{IEEEtran}
\bibliography{quantum}
\balance

\end{document}